# Triboelectric Junction:
# A Model for Dynamic Metal-Semiconductor Contacts

**Xiaote Xu[1,2], Zhong Lin Wang[3,4]\*, Zhengbao Yang[1,2]\***

**1 Department of Mechanical and Aerospace Engineering, Hong Kong University of Science and Technology, Clear Water Bay, Hong Kong, China**
**2 Department of Mechanical Engineering, City University of Hong Kong, Hong Kong SAR, China**
**3 Beijing Institute of Nanoenergy and Nanosystems, Chinese Academy of Sciences, Beijing, 101400, People's Republic of China**
**4 School of Materials Science and Engineering, Georgia Institute of Technology, Atlanta, GA, 30332, USA**

**Email:** zbyang@ust.hk; zhong.wang@mse.gatech.edu

## Abstract

Static metal-semiconductor contacts are classified into Ohmic contacts and Schottky contacts. As for dynamic metal-semiconductor contacts, the in-depth mechanism remains to be studied. We here define a "triboelectric junction" model for analyzing dynamic metal-semiconductor contacts, where a space charge region induced by the triboelectric effect dominates the electron-hole separation process. Through theoretical analysis and experiments, we conclude that the triboelectric junction affects the electric output in two aspects: 1) the junction direction determines the output polarity; 2) the junction strength determines the output amplitude. The junction direction and junction strength are both related to the electron-affinity difference between the contact metal and semiconductor. We find that the standard electrode potential in electrochemistry best describes the electron affinity of a dynamic metal-semiconductor contact.

**Keywords:** Schottky; Semiconductor; Tribology; Interface; Dynamic contact; Generator; Energy harvesting.

## Introduction

Metal-semiconductor contacts, including Ohmic contacts and Schottky contacts, are fundamental for modern electronics[1]. Ohmic contacts show linear current-voltage curves, presenting non-rectifying characteristics. Ohmic contacts are usually desired for effective charge conduction between semiconductor and external circuitry, such as source/drain-semiconductor contacts in transistors[2,3]. Schottky contacts, in which Schottky barriers form at metal-semiconductor interfaces, show rectifying characteristics[4]. Schottky contacts are used in different scenarios, such as rectification in diodes[5] and electron-hole separation in solar cells[6]. Overall, metal-semiconductor contacts significantly contribute to the advancement of both the electronic and energy industry.

Classical metal-semiconductor contacts discussed above are static contacts. Recently, dynamic metal-semiconductor contacts have started to gain attention in the energy-harvesting field[7–12]. However, dark clouds are hanging over the mechanism of dynamic metal-semiconductor contacts[13,14]. Schottky junction theory, established for static contacts, is widely adopted for explaining dynamic metal-semiconductor contacts[13,14]. However, our experiments show some counterexamples:

1) Al and Ag have similar work-function values of about 4.28 eV and 4.26 eV, respectively, which form similar Schottky barrier heights when contacting the same silicon. However, the output polarity of a dynamic Al-Si contact is opposite to that of a dynamic Ag-Si contact (see Fig. S1e);

2) The output polarity remains the same when sliding the same metal material on silicon wafers with different Fermi levels. However, according to the Schottky junction theory, those Ohmic contacts should not generate electric outputs, especially when using heavily doped silicon (discussed later in Fig. 3f and Fig. 3g).

The counterexamples indicate that the classical Schottky junction theory (static contacts) does not work in dynamic metal-semiconductor contacts. In classical semiconductor physics, the surface state is an important factor in the analysis of contact issue[15], not to mention the unavoidable triboelectric effect during dynamic contacts. Previous studies have also proven that the triboelectric effect plays a dominant role in the electrical response of dynamic semiconductor-semiconductor contacts[16–18]. Thus, we need a new model for dynamic metal-semiconductor contacts.

We here define a "triboelectric junction" model for analyzing dynamic metal-semiconductor contacts. We argue that the triboelectric junction dominates the electron-hole separation process in dynamic contacts. We present a theoretical analysis of the triboelectric junction, elucidating the electric field distribution and electric potential distribution. A material's figure-of-merit is developed to link the output voltage to the intrinsic material properties (electron affinity). This work provides a new perspective on the mechanism of dynamic metal-semiconductor contacts, which might further trigger new fundamental discoveries and applications.

## Results

### Theory of the triboelectric junction

Let's start by learning the photovoltaic effect, which uses semiconductors to convert light into electricity. As a mature technology, the working mechanism of a solar cell is well-established[1,19], which can be interpreted from the following steps (see Fig. 1a and Fig. S2a): 1) Formation of a PN junction/Schottky junction at the contact interface due to the alignment of Fermi levels; 2) Nonequilibrium electron-hole generation excited by photons with energy greater than semiconductor bandgap; 3) Electron-hole separation driven by the PN junction/Schottky junction, resulting in direct-current output. A junction region, also known as a space charge region/depletion region, is consisted of immobile positive and negative ions, acting as the electric field for driving electron-hole separation.

Analogous to the PN junction/Schottky junction (space charge region) in static contacts, we define a triboelectric junction for interpreting the electric response in dynamic metal-semiconductor contacts (see Fig. 1b). We define the triboelectric junction as a space charge region induced by the triboelectric effect, which only exists in dynamic contacts. Taking the dynamic Al-Si contact as an example (Electron affinity: Al < Si), Al tends to donate electrons, and Si tends to accept electrons during a dynamic contact, leaving positive ions on the Al surface and negative ions on the Si surface. Thus, a triboelectric junction is formed, with an electric field direction from Al to Si (see Fig. 1d). Note that the space charge region in metals is extremely thin (almost non-existent). Ions in metals are drawn in the schematic diagram for easy understanding.

Overall, the working mechanism of a dynamic semiconductor generator is summarized as follows: 1) During a dynamic contact, two necessary processes coincide, which are the formation of the triboelectric junction at the contact interface (see Fig. 1d) and the generation of nonequilibrium electron-hole pairs in the semiconductor (see Fig. 1e), respectively; 2) Nonequilibrium electron-hole pairs are further separated by the triboelectric junction (see Fig. 1f), resulting in direct-current output. The triboelectric effect results in the formation of a triboelectric junction through electron transfer at the interface, as well as the generation of nonequilibrium electron-hole pairs through electron excitation in the semiconductor.

The working mechanism of a dynamic semiconductor generator is similar to a solar cell, in which the junction region (space charge region) drives electron-hole separation to generate direct-current output. However, there are some significant differences: 1) Mechanism of the junction formation: The triboelectric junction in a dynamic semiconductor generator is formed by the triboelectric effect, which only exists in dynamic contacts, while the PN junction/Schottky junction (static contacts) in a solar cell is formed due to the alignment of Fermi levels. 2) Mechanism of the nonequilibrium electron-hole generation: Photons with energy greater than the semiconductor bandgap excite the electron-hole generation in a solar cell; however, the excitation energy for the electron-hole generation in a dynamic semiconductor generator is from the contact-electrification, likely the released energy from the formation of a chemical bond[11,14].

We argue that the triboelectric junction, rather than the Schottky junction, dominates the electron-hole separation process in dynamic metal-semiconductor contacts. Both types of junctions involve a space charge region. However, the Schottky junction is formed due to the alignment of Fermi levels, while the triboelectric junction is formed by the much stronger triboelectric effect[16–18]. Moreover, the triboelectric junction also interferes with the dynamic equilibrium of charge movement necessary for maintaining an intrinsic Schottky junction, that is, the triboelectric effect induced by sliding affects the structure of the energy-band diagram[14]. From the perspective of the "triboelectric junction" model, such a complex coupling effect can be simplified as the influence of the Fermi level on the triboelectric junction.

**Theoretical analysis of the triboelectric junction**

A triboelectric junction is a space charge region induced by the triboelectric effect, exhibiting characteristics similar to an abrupt heterojunction. In an abrupt heterojunction, the ion types in the space charge region change abruptly from positive ions to negative ions across the contact interface. In a triboelectric junction, one material donates electrons while the other material accepts electrons, which also presents an abrupt change in ion type. Thus, the analysis of the triboelectric junction can be analogous to the abrupt heterojunction[1,20]. To simplify the analysis, we discuss the triboelectric junction in a thermal equilibrium condition. In a triboelectric junction, the material that accepts (donates) electrons forms a negative (positive) ion zone with an ion concentration of $N_A$ ($N_D$) (see Fig. 2a). The widths of the negative and positive ion zone are defined as $x_A$ and $x_D$, respectively. The contact interface of the negative and positive ion zone is defined as $x=0$. Thus, we obtain the charge density distribution, as shown in Fig. 2b:

$$\rho(x) = \begin{cases} -qN_A \, , & -x_A \le x \le 0 \\ qN_D \, , & 0 \le x \le x_D \end{cases} . \tag{1}$$

According to the electron transfer model of the triboelectric effect[21], the space charge amount (the triboelectric charge amount) in the negative ion zone is equal to that of the positive ion zone:

$$qN_A x_A = qN_D x_D = Q \, , \tag{2}$$

where $Q$ is the space charge amount (the triboelectric charge amount) per unit area.

According to the Poisson equation[1,20], we obtain

$$-\frac{d^2V(x)}{dx^2} = \frac{dE(x)}{dx} = \frac{\rho(x)}{\varepsilon(x)} = \begin{cases} -\dfrac{qN_A}{\varepsilon_A} \, , & -x_A \le x \le 0 \\ \dfrac{qN_D}{\varepsilon_D} \, , & 0 \le x \le x_D \end{cases} . \tag{3}$$

Integrating the above equations gives the electric field distribution, as shown in Fig. 2c:

$$E(x) = -\frac{dV(x)}{dx} = \begin{cases} -\dfrac{qN_A(x+x_A)}{\varepsilon_A} \, , & -x_A \le x \le 0 \\ \dfrac{qN_D(x-x_D)}{\varepsilon_D} \, , & 0 \le x \le x_D \end{cases} . \tag{4}$$

At $x=0$, the electric field strength reaches its maximum for the negative ion zone ($E_{Am}$) and positive ion zone ($E_{Dm}$), respectively. Thus, we obtain

$$E_{Am} = -\frac{qN_A x_A}{\varepsilon_A} = -\frac{Q}{\varepsilon_A} \, , \tag{5}$$

$$E_{Dm} = -\frac{qN_D x_D}{\varepsilon_D} = -\frac{Q}{\varepsilon_D}. \tag{6}$$

Note that the electric field distribution is discontinuous across the contact interface ($x$=0) due to the change in the dielectric permittivity. The electric field direction and electric field strength of the triboelectric junction are named junction direction and junction strength, respectively, for convenience.

Integrating Equation 4, we obtain the electric potential distribution, as shown in Fig. 2d:

$$V(x) = \begin{cases} \dfrac{qN_A(x+x_A)^2}{2\varepsilon_A}, & -x_A \leq x \leq 0 \\ \dfrac{qN_D}{\varepsilon_D}(x_D - \dfrac{x}{2})x + \dfrac{qN_A x_A^2}{2\varepsilon_A}, & 0 \leq x \leq x_D \end{cases}. \tag{7}$$

Thus, we obtain the electric potential across the triboelectric junction, namely the junction voltage:

$$V_{TJ} = \frac{qN_A x_A^2}{2\varepsilon_A} + \frac{qN_D x_D^2}{2\varepsilon_D} = \frac{Q^2}{2\varepsilon_A qN_A} + \frac{Q^2}{2\varepsilon_D qN_D}. \tag{8}$$

Note that the theoretical analysis above is the general model of the triboelectric junction. A more specific derivation is shown in Note S1.

The general model of the triboelectric junction can be applied to both dynamic semiconductor-semiconductor contacts and metal-semiconductor contacts. For dynamic semiconductor-semiconductor contacts using different contact materials, different dielectric-permittivity values are adopted, similar to the abrupt heterojunction. For dynamic semiconductor-semiconductor contacts using the same contact material but with different Fermi levels, the same dielectric-permittivity value is adopted, similar to the abrupt PN junction. For dynamic metal-semiconductor contacts, the junction region is considered to exist only in the semiconductor ($\varepsilon_m=\infty$), similar to the one-sided abrupt PN junction (P$^+$N junction or PN$^+$ junction). For dynamic metal-metal contacts, the junction strength is negligibly small, considering that the dielectric permittivity of metals is infinite ($\varepsilon_m=\infty$). Dynamic contacts with insulators are not discussed in the triboelectric junction model.

For metal-semiconductor contacts, the triboelectric junction can be divided into two categories and further simplified: 1) The semiconductor accepts electrons (see Fig. 2e, Fig. 2f, and Fig. S3); and 2) The semiconductor donates electrons (see Fig. 2g, Fig. 2h and Fig. S3). Equation 8 reduces to

$$V_{TJ} = \frac{Q^2}{2\varepsilon_s qN}, \tag{9}$$

where $N$ is $N_A$ or $N_D$, depending on whether the semiconductor accepts or donates electrons.

The junction voltage analyzed in a thermal equilibrium condition is the theoretical maximum junction voltage. Note that the open-circuit voltage is closely related to, but not equal to, the junction voltage. In the photovoltaic effect, the junction voltage (built-in voltage) is regarded as the upper limit of the open-circuit voltage[22,23]. For a solar cell operated in an open-circuit state, a potential difference, caused by the accumulation of charge carriers at electrodes, will cancel out the junction voltage to a certain degree[24,25]. For a dynamic semiconductor generator, the nonequilibrium electron-hole generation in the semiconductor is necessary, which is not actually in a thermal equilibrium condition. The open-circuit voltage would never reach the junction voltage due to the accumulation of the charge carriers at electrodes, similar to the photovoltaic effect. However, a larger junction voltage does contribute to a larger open-circuit voltage.

**Material's figure-of-merit of the triboelectric junction**

From Equations 5,6,8, and 9, we know that both the junction strength and junction voltage are positively correlated to the $Q$ (the triboelectric charge amount per unit area). The triboelectric effect is a complex phenomenon[21] that, so far, no equation can accurately express the generation of triboelectric charge. However, we can conclude that the $Q$ is related to the contact materials[26,27] ($k_M$), mechanical input[16] ($k_I$), and environmental factor[28,29] ($k_E$). Thus, we obtain

$$Q = f(k_M, k_I, k_E) \ . \tag{10}$$

We concern more about the intrinsic material properties $k_M$, which determines the tendency of donating/accepting electrons. The triboelectric series is usually utilized to analyze the triboelectric effect. Although intensive efforts have been made to extend the triboelectric series[26,27], most of the materials presented in the triboelectric series are insulators, in which the triboelectric charge can be effectively stored on the surface after contact, and measured by electrostatic induction. As for the triboelectric junction, most of the contact materials (metals and semiconductors) are not presented in the triboelectric series. We thus need a new method to quantify the electric outputs of dynamic semiconductor generators.

Theoretically, the tendency of donating/accepting electrons is determined by the electron affinity of the contact materials[30–33]. Thus, the electron-affinity values of metal and semiconductors contribute to understanding the triboelectric junction. Electron affinity has also been researched in chemistry and semiconductor physics. However, the electron-affinity values in chemistry and semiconductor physics are characterized in a gaseous state and at a semiconductor-vacuum interface, respectively, which cannot be directly applied to the triboelectric effect in a solid-solid contact. The interatomic force in a solid-solid contact is much stronger than that of a gaseous state or semiconductor-vacuum interface. In electrochemistry, a reactivity series of different metals, characterized by standard electrode potential (SEP)[34], is adopted to analyze the tendency of donating/accepting electrons between metals and other metal ions in solution. In our experiments, we find that the standard electrode potential can be used as a reference for electron affinity in the analysis of the triboelectric junction. For example, electrochemical displacement plating of metal (Cu, Ag, Pt, and Au) on Silicon, namely junction delineation, is developed as a technology to examine junction depth in the semiconductor industry. In the junction delineation process, silicon donates electrons, and metal ions in solution accept electrons[35,36], which is consistent with our experiment results that silicon donates electrons and metal (Cu, Ag, Pt, and Au) accepts electrons during dynamic contact (discuss later in Fig. 4d, Fig. 4e and Fig. 4g). The possible reason is attributed to the chemical reaction accompanied by tribology, namely mechanochemistry[37–39].

We introduce a material's figure-of-merit[40]:

$$k_M = \frac{\chi_M - \chi_S}{|\chi_S|}, \tag{11}$$

where $\chi_M$ and $\chi_S$ are the electron affinities of the contact metal and semiconductor, respectively. $k_M$ is a dimensionless parameter for characterizing the relative relationship of the electron-affinity values of the contact materials, which ultimately affects the electric output. Interpreted from $k_M$, the sign of $k_M$ reflects the junction direction; the absolute value of $k_M$ reflects the junction strength.

**The junction direction determines the output polarity**

The total junction at a dynamic interface has two components: one is the triboelectric junction; the other is contributed by the intrinsic Fermi levels of the two contact materials. The triboelectric junction is a space charge region that dominates the electron-hole separation process. The junction direction determines the output polarity; the junction strength determines the output amplitude. Note that the "output" here refers to the open-circuit voltage for convenience. Dynamic metal-silicon contacts were adopted to verify the model, which exhibits several advantages as follows: 1) Metal and silicon are both single-element materials, whose electron-affinity values can refer to the well-established standard electron potential[34] (see Fig. S4); 2) The triboelectric junction is considered to exist only in silicon since the dielectric permittivity of metal is infinite ($\varepsilon_m = \infty$); 3) Nonequilibrium electron-hole generation only occurs in the silicon.

We selected three representative materials (Electron affinity: Al<Si<Cu, see Fig. 3b) to examine the effect of junction direction on the output polarity. The junction direction of a dynamic Al-Si contact is from Al to Si (see Fig. 3d). In contrast, it is from Si to Cu for dynamic a Cu-Si contact (see Fig. 3e). The dynamic Al-Si contact generates a negative output (see Fig. 3f). In contrast, the dynamic Cu-Si contact generates a positive output (see Fig. 3g). The output polarity is determined by the junction direction. The output polarity was further verified by adopting oxide-free Al for dynamic contacts in a helium glovebox, showing the same negative polarity as untreated Al in ambient (see Fig. S5a). Note that the thicknesses of the natural oxide layers of Al[41–43] and silicon[44,45] are only about several nanometers, which are easily destroyed in sliding contacts and thus neglected.

Furthermore, the output polarity remains the same when sliding the same metal material on silicon wafers with different Fermi-levels, *i.e.*, heavily doped N-type silicon wafer (N$^+$ Si), N-type silicon wafer (N Si), intrinsic silicon wafer (I Si), P-type silicon wafer (P Si) and heavily doped P-type silicon wafer (P$^+$ Si). Silicon wafers with different Fermi levels are fabricated by doping processes with different impurity types and concentrations[46–48]. The dopants only take a small portion compared to the main body of silicon. A dopant concentration above 0.1% can be regarded as heavily doped for silicon wafers. Thus, the main material atoms that contact each other at the interface are still silicon atoms and metal atoms. Such a phenomenon is strong evidence for the triboelectric junction.

We tested more metal materials to verify the triboelectric junction model. For metals with electron-affinity values smaller than Si, such as Mg, Al, Zn, and Sn, negative outputs are generated. For metals with electron-affinity values larger than Si, such as Cu, Ag, Pt, and Au, positive outputs are generated (see Fig. 3h). Interpreted from the material's figure-of-merit $k_M$, the output polarity is negative when $\chi_M$-$\chi_S$ < 0, while it is positive when $\chi_M$-$\chi_S$ > 0. In addition to the dynamic metal-semiconductor contacts, the triboelectric junction is also verified in a dynamic semiconductor-semiconductor contact (see Fig. S5b).

**The junction strength determines the output amplitude**

Junction voltage is the integral of the electric field across the junction region. Thus, a larger junction strength (electric field strength) contributes to a larger junction voltage, resulting in a larger open-circuit voltage. "$E_{Mg-}$" in Fig. 4b represents the electric field of the triboelectric junction in a dynamic Mg-Semiconductor contact, which generates negative output. The symbols of other electric fields and open-circuit voltages share the same naming rules. For negative output, Mg is easier to donate electrons than Sn when dynamically contacting with silicon, resulting in a larger junction strength of a dynamic Mg-Si contact than that of a dynamic Sn-Si contact (junction strength: $E_{Mg-}$>$E_{Sn-}$, see Fig. 4b and Fig. 4c). Accordingly, the output amplitude of a dynamic Mg-Si contact is larger than that of a dynamic Sn-Si contact. Further comparing different metals, we observe the following regularity of output amplitude: $|V_{Mg-}|$>$|V_{Al-}|$>$|V_{Zn-}|$>$|V_{Sn-}|$ (see Fig. 4f). Similarly, for positive output, Au is easier to accept electrons than Cu when dynamically contacting with silicon, resulting in a larger junction strength of a dynamic Au-Si contact than that of a dynamic Cu-Si contact (junction strength: $E_{Au+}$>$E_{Cu+}$, see Fig. 4d and Fig. 4e). Accordingly, the output amplitude of a dynamic Au-Si contact is larger than that of a dynamic Cu-Si contact. Further comparing different metals, we observe the following regularity of output amplitude: $|V_{Au+}|$>$|V_{Pt+}|$>$|V_{Ag+}|$ >$|V_{Cu+}|$ (see Fig. 4g). However, output instability was observed in dynamic metal-silicon contact (see Fig. S7), resulting in unstable repeatability on the output amplitude.

We further examined the triboelectric junction model using a semiconductor polymer, Poly (3, 4-ethylenedioxythiophene):poly (styrene sulfonate) (PEDOT:PSS)[10–12], which generates a stable output amplitude. For negative output, the output amplitude ($|V_{Mg-}|>|V_{Al-}|>|V_{Zn-}|>|V_{Sn-}|$) shows a strong correlation with the junction strength ($|E_{Mg-}|>|E_{Al-}|>|E_{Zn-}|>|E_{Sn-}|$), see Fig. 4h). For positive output, we compared a dynamic Si-Al contact and a dynamic PEDOT:PSS-Al contact, both with the Al components grounded. The output amplitude ($V_{PEDOT:PSS+}>V_{Si+}$) also shows a strong correlation with the junction strength ($E_{PEDOT:PSS+}>E_{Si+}$, see Fig. 4i). Interpreted from the material's figure-of-merit $k_M$, a larger $|\chi_M - \chi_S|$ contributes to a larger output amplitude.

## Discussion

Compared with the mature theory for static contacts, the theory for dynamic metal-semiconductor contacts is in its infancy. The triboelectric junction model developed here is simplified, and some challenges are left to be addressed in the future.

(1) Quantifying the released energy from the triboelectric effect. The "bindington", an energy quantum released from the formation of a chemical bond during contact-electrification, is considered more likely to be the excitation energy for nonequilibrium electron-hole generation in semiconductor[11,14]. The released energy from dynamic contact may vary in different contact pairs. Those released energy greater than the semiconductor bandgap can excite electron-hole generation. Despite its importance, quantifying the released energy from the triboelectric effect remains challenging due to its complexity.

(2) Formulating current-voltage characteristics. Compared to the PN junction/Schottky junction, the current-voltage characteristic of the triboelectric junction, which possesses multiple combinations of Fermi levels, is more complex. In deviation of the ideal current-voltage characteristics of a PN junction, namely Shockley equation[19], the following equation $qV=E_{Fn}-E_{Fp}$ links the voltage with quasi-Fermi levels, where $q$ is the electric charge of an electron, $V$ is the applied voltage, $E_{Fn}$ and $E_{Fp}$ are the quasi-Fermi levels for the n-type region and p-type region, respectively. The Boltzmann relation connects the quasi-Fermi levels ($E_{Fn}$ and $E_{Fp}$) with the charge carrier densities (electron and hole density), which ultimately determine the current output. Thus, the ideal current-voltage characteristics of a PN junction can be derived[1,19]. Since the triboelectric effect is difficult to quantitatively formulate, linking the triboelectric effect and semiconductor physics to formulate the current-voltage characteristic for the triboelectric junction remains a significant challenge.

(3) Formulating electro-mechanical responses. In the photovoltaic effect, the electric output under different light intensities can be analyzed based on the current-voltage characteristic[1]. Similarly, for certain contact materials, the electro-mechanical response of a dynamic semiconductor generator can be analyzed by combining the mechanical input $K_I$ (see Equation 10) and the current-voltage characteristic of the triboelectric junction.

## Materials and methods

### Device Fabrication and Electrical Characterization

All single-polished silicon wafers (Zhejiang Lijing Silicon Material Co., Ltd., China) have the same crystal plane orientation [100] and thickness (500 μm). The resistances of $N^+/P^+$ silicon, N/P silicon, and I silicon are 0.001-0.005 Ω·cm,1-20 Ω·cm, and >5000 Ω·cm, respectively. A gold film is sputtered on the unpolished side of the silicon wafer as a bottom electrode. The silicon wafer was tailored into a rectangle (of size 20 mm × 30 mm) for device fabrication. The rectangular silicon, equipped with an outgoing line on the bottom gold electrode, was attached to a PVC substrate. Different metal foils (Mg, Al, Zn, Fe, Sn, Cu, Ag, Pt, and Au) were tailored into a square (of size 5 mm × 5 mm), equipped with an outgoing line on the back side, and then attached to PVC substrates. PEDOT:PSS solution（Shanghai Ouyi Organic Photoelectric Material, OE-001）was drop-casted into a cleanroom wiper (of size 20 mm × 30 mm) to form a PEDOT:PSS-textile composite and then dried in a 70 °C oven for 60 min. The PEDOT:PSS-textile composite was attached to a copper bottom electrode (Benyida Company, thickness 50 μm) on a PVC substrate. The open-circuit voltage was characterized by an oscilloscope (Rohde and Schwarzrte, RTE1024) under a normal force of 2 N. If not specified, the semiconductor component is grounded. The output voltage of the intrinsic silicon-based device was further processed to eliminate the baseline (usually tens of millivolts) induced by the photovoltaic effect (see Fig. S9).

**Figures**

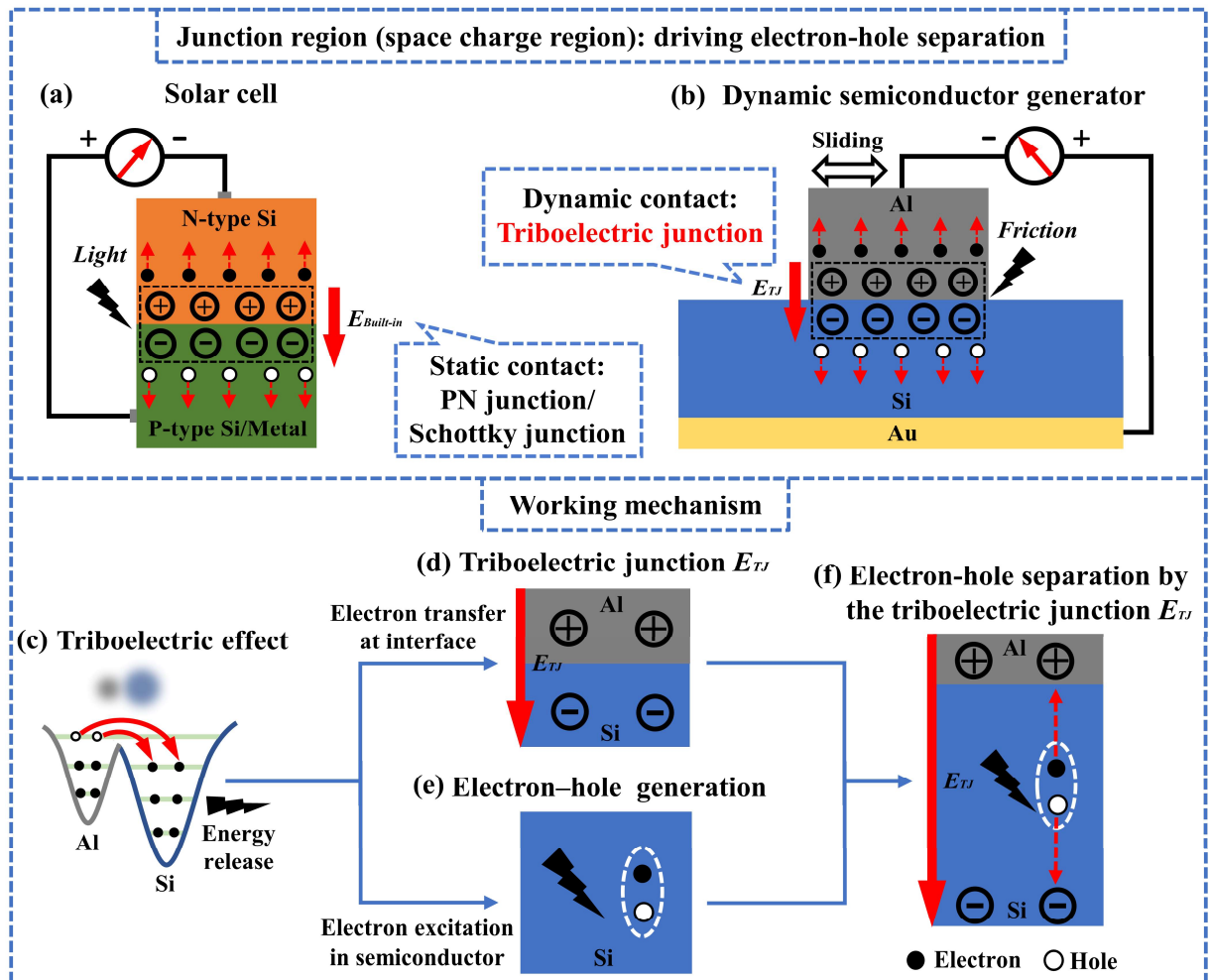

**Fig. 1. Theory of the triboelectric junction.** (a) A PN junction/Schottky junction (static contact) drives the electron-hole separation in a solar cell. Note that holes in the metal should be neglected for the Schottky junction in the schematic diagram; (b) A triboelectric junction (dynamic contact) dominates the electron-hole separation process in a dynamic semiconductor generator; (c) The electron-cloud-potential-well model of the triboelectric effect in a dynamic Al-Si contact, including the electron transfer process between the contact materials and the related energy release; (d) Triboelectric junction, a space charge region induced by the triboelectric effect (electron transfer at the contact interface); (e) Nonequilibrium electron-hole generation in semiconductor, excited by the released energy from the triboelectric effect; (f) Electron-hole separation driven by the triboelectric junction $E_{TJ}$.

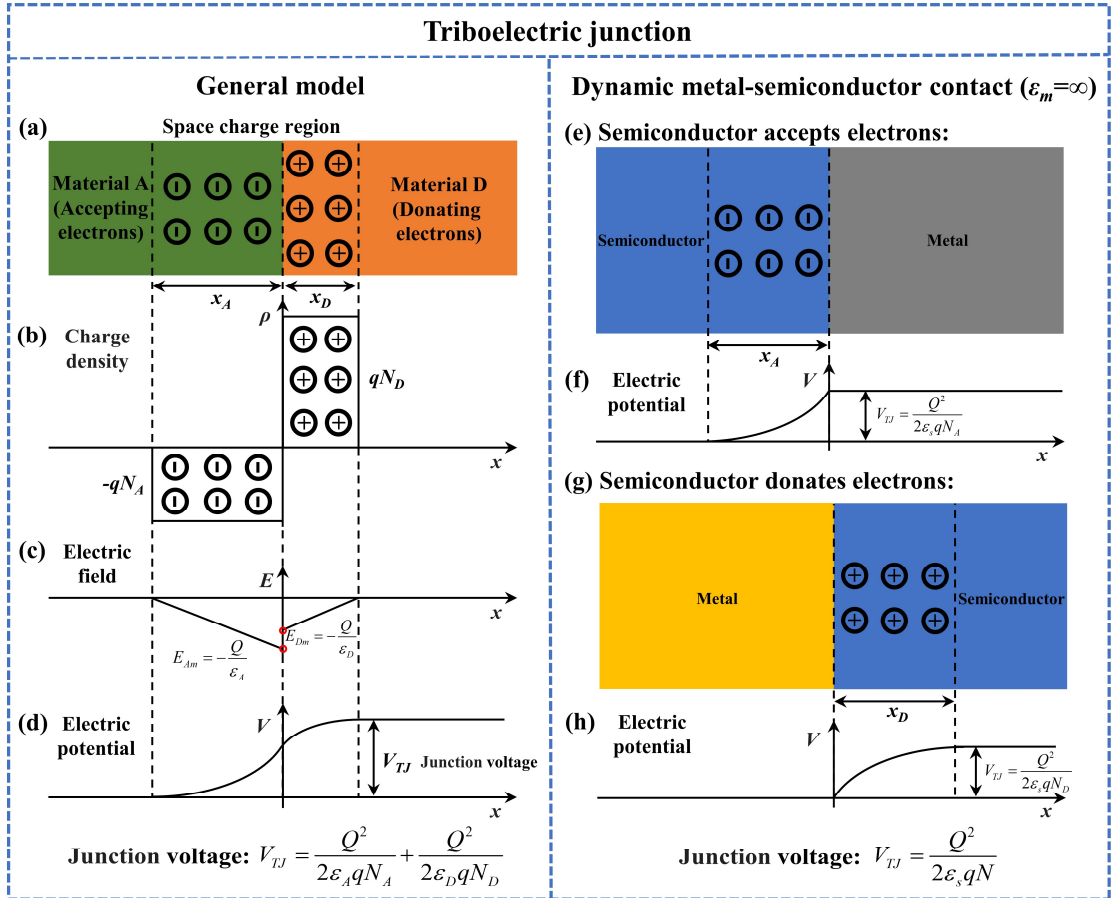

**Fig. 2. Theoretical analysis of the triboelectric junction.** (a) The general model of the triboelectric junction; and the corresponding (b) Charge density distribution; (c) Electric field distribution (in a case that $\varepsilon_A < \varepsilon_D$); (d) Electric potential distribution; (e) Model and (f) Electric potential distribution of dynamic metal-semiconductor contacts that the semiconductor accepts electrons; (g) Model and (h) Electric potential distribution of dynamic metal-semiconductor contacts that the semiconductor donates electrons.

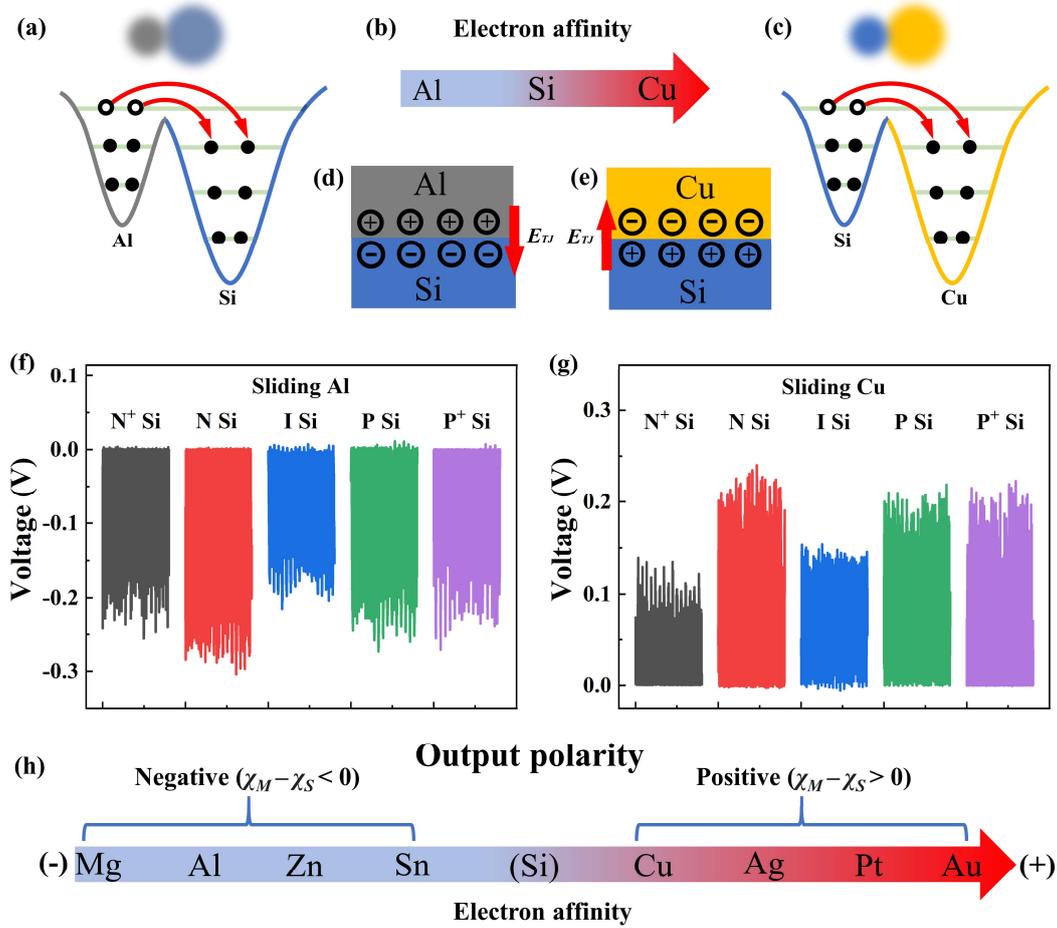

**Fig. 3. The junction direction determines the output polarity.** (a) Schematic diagram of the triboelectric effect in a dynamic Al-Si contact; (b) Electron affinity; (c) Dynamic Cu-Si contact; (d) The triboelectric junction in a dynamic Al-Si contact and (e) Dynamic Cu-Si contact; (f) Dynamic Al-Si contacts generate negative outputs; (g) Dynamic Cu-Si contacts generate positive outputs; (h) The output polarity is determined by the junction direction (the sign of $\chi_M$-$\chi_S$).

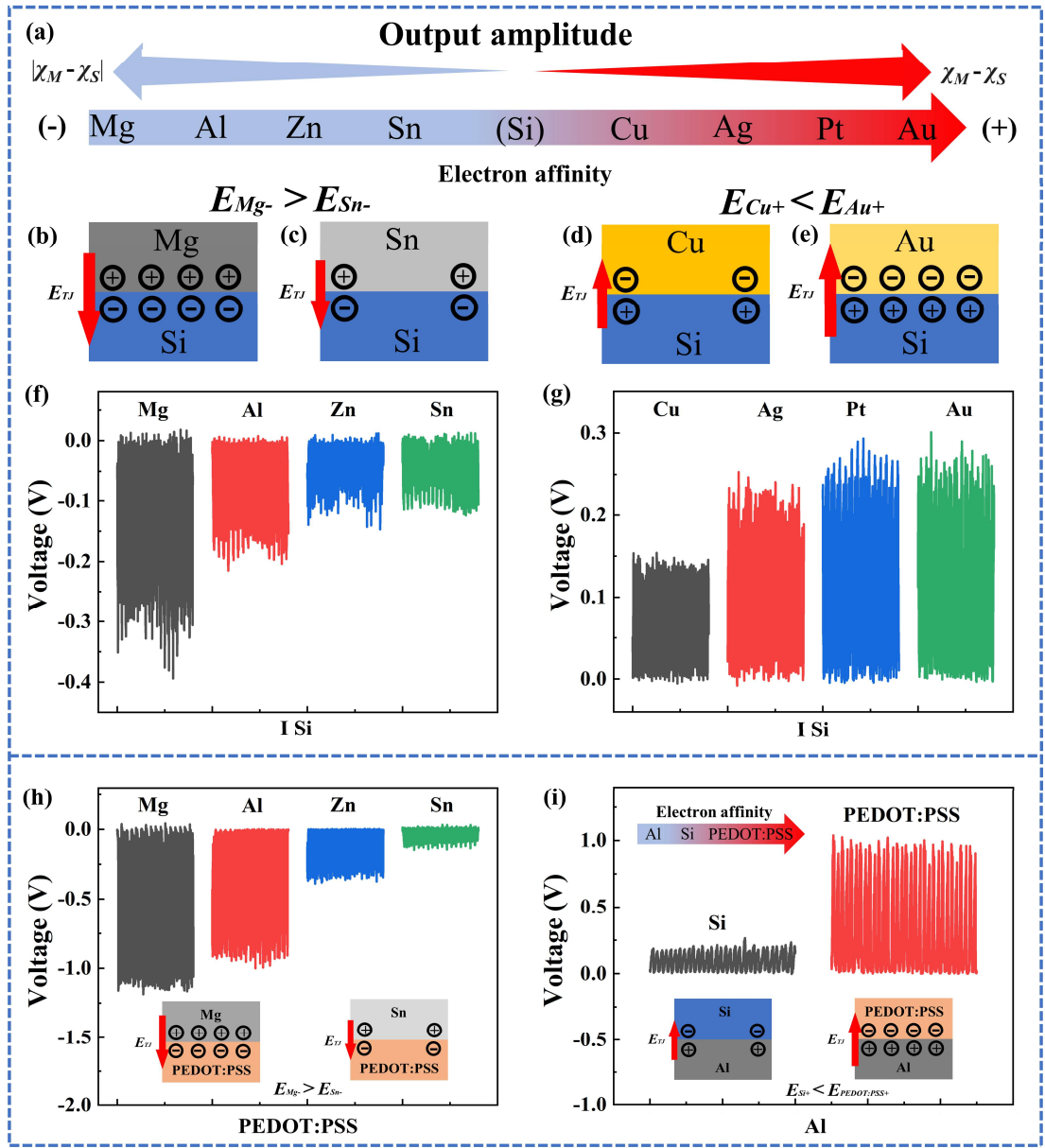

**Fig. 4. The junction strength determines the output amplitude.** (a) For dynamic metal-semiconductor contacts, a larger junction strength ($|\chi_M-\chi_S|$) results in a larger output amplitude; (b) The junction strength of a dynamic Mg-Si contact is larger than that of (c) Dynamic Sn-Si contact; (d) The junction strength of a dynamic Cu-Si contact is smaller than that of (d) Dynamic Au-Si contact; (f) Negative output amplitudes for dynamic metal-silicon contacts ($V_{Mg-}>V_{Al-}>V_{Zn-}>V_{Sn-}$); (g) Positive output amplitudes for dynamic metal-silicon contacts ($V_{Au+}>V_{Pt+}>V_{Ag+}>V_{Cu+}$); (h) Negative output amplitudes for dynamic metal-PEDOT:PSS contacts ($V_{Mg-}>V_{Al-}>V_{Zn-}>V_{Sn-}$); (i) Positive output amplitudes for a dynamic Si-Al contact and a dynamic PEDOT:PSS-Al contact ($V_{PEDOT:PSS+}>V_{Si+}$).

**Funding:** The work described in this paper was supported by General Research Grant (Project Nos. 11212021; 11210822) from the Research Grants Council of the Hong Kong Special Administrative Region.

**Author contributions:** ZY and XX conceived the project and designed the studies. XX performed experiments and analyzed the experimental data. ZY and XX wrote the manuscript. ZY and ZLW supervised the project.

**Competing interests:** The authors declare no competing interests.

**Data and materials availability:** All data needed to evaluate the conclusions in this paper are present in the paper and/or the Supplementary Information. Additional data and raw data are available upon request from the authors.

# Reference

1. Sze, S. M. & Ng, K. K. *Physics of Semiconductor Devices. Wiley* (2021).
2. Wang, Y. *et al.* Van der Waals contacts between three-dimensional metals and two-dimensional semiconductors. *Nature* **568**, 70–74 (2019).
3. Shen, P. C. *et al.* Ultralow contact resistance between semimetal and monolayer semiconductors. *Nature* **593**, 211–217 (2021).
4. Schottky, W. Halbleitertheorie der Sperrschicht. *Naturwissenschaften* **26**, 843 (1938).
5. Harada, T., Ito, S. & Tsukazaki, A. Electric dipole effect in PdCoO2/β-Ga2O3 Schottky diodes for high-temperature operation. *Sci. Adv.* **5**, 1–8 (2019).
6. Went, C. M. *et al.* A new metal transfer process for van der Waals contacts to vertical Schottky-junction transition metal dichalcogenide photovoltaics. *Sci. Adv.* **5**, (2019).
7. Zhang, Z. *et al.* Tribovoltaic Effect on Metal–Semiconductor Interface for Direct-Current Low-Impedance Triboelectric Nanogenerators. *Adv. Energy Mater.* **10**, 1–8 (2020).
8. Liu, J. *et al.* Direct-current triboelectricity generation by a sliding Schottky nanocontact on MoS2 multilayers. *Nat. Nanotechnol.* **13**, 112–116 (2018).
9. Shao, H., Fang, J., Wang, H., Dai, L. & Lin, T. Polymer-Metal Schottky Contact with Direct-Current Outputs. *Adv. Mater.* **28**, 1461–1466 (2016).
10. Yang, R., Benner, M., Guo, Z., Zhou, C. & Liu, J. High-Performance Flexible Schottky DC Generator via Metal/Conducting Polymer Sliding Contacts. *Adv. Funct. Mater.* **31**, 1–12 (2021).
11. Meng, J. *et al.* Durable flexible direct current generation through the tribovoltaic effect in contact-separation mode. *Energy Environ. Sci.* **28**, 5159–5167 (2022).
12. Meng, J. *et al.* Flexible Textile Direct-Current Generator Based on the Tribovoltaic Effect at Dynamic Metal-Semiconducting Polymer Interfaces. *ACS Energy Lett.* **6**, 2442–2450 (2021).
13. Yang, R. *et al.* Semiconductor-based dynamic heterojunctions as an emerging strategy for high direct-current mechanical energy harvesting. *Nano Energy* **83**, 105849 (2021).
14. Lin, S. & Lin Wang, Z. The tribovoltaic effect. *Mater. Today* **62**, 111–128 (2023).
15. Cowley, A. M. & Sze, S. M. Surface states and barrier height of metal-semiconductor systems. *J. Appl. Phys.* **36**, 3212–3220 (1965).

16. Chen, Y. *et al.* Friction-Dominated Carrier Excitation and Transport Mechanism for GaN-Based Direct-Current Triboelectric Nanogenerators. *ACS Appl. Mater. Interfaces* **14**, 24020–24027 (2022).

17. Zhang, Z. *et al.* Semiconductor Contact-Electrification-Dominated Tribovoltaic Effect for Ultrahigh Power Generation. *Adv. Mater.* **34**, 2200146 (2022).

18. Wang, Z. *et al.* Achieving an ultrahigh direct-current voltage of 130 V by semiconductor heterojunction power generation based on the tribovoltaic effect. *Energy Environ. Sci.* **15**, 2366–2373 (2022).

19. Landsberg, P. T. An introduction to the theory of photovoltaic cells. *Solid State Electron.* **18**, 1043–1052 (1975).

20. Shockley, W. The Theory of p-n Junctions in Semiconductors and p-n Junction Transistors. *Bell Syst. Tech. J.* **28**, 435–489 (1949).

21. Wang, Z. L. & Wang, A. C. On the origin of contact-electrification. *Mater. Today* **30**, 34–51 (2019).

22. He, Z. *et al.* Simultaneous enhancement of open-circuit voltage, short-circuit current density, and fill factor in polymer solar cells. *Adv. Mater.* **23**, 4636–4643 (2011).

23. Luo, J. *et al.* Enhanced open-circuit voltage in polymer solar cells. *Appl. Phys. Lett.* **95**, (2009).

24. Qi, B. & Wang, J. Open-circuit voltage in organic solar cells. *J. Mater. Chem.* **22**, 24315–24325 (2012).

25. Elumalai, N. K. & Uddin, A. Open circuit voltage of organic solar cells: An in-depth review. *Energy Environ. Sci.* **9**, 391–410 (2016).

26. Zou, H. *et al.* Quantifying the triboelectric series. *Nat. Commun.* **10**, 1–9 (2019).

27. Zou, H. *et al.* Quantifying and understanding the triboelectric series of inorganic non-metallic materials. *Nat. Commun.* **11**, 1–7 (2020).

28. Wang, Z. *et al.* A humidity-enhanced silicon-based semiconductor tribovoltaic direct-current nanogenerator. *J. Mater. Chem. A* **10**, 25230–25237 (2022).

29. Xu, C. *et al.* Raising the Working Temperature of a Triboelectric Nanogenerator by Quenching Down Electron Thermionic Emission in Contact-Electrification. *Adv. Mater.* **30**, 201803968 (2018).

30. Chen, G., Au, C. & Chen, J. Textile Triboelectric Nanogenerators for Wearable Pulse Wave Monitoring. *Trends Biotechnol.* **39**, 1078–1092 (2021).

31. Xiao, X., Chen, G., Libanori, A. & Chen, J. Wearable Triboelectric Nanogenerators for Therapeutics. *Trends Chem.* **3**, 279–290 (2021).

32. Chen, A., Zhang, C., Zhu, G. & Wang, Z. L. Polymer Materials for High-Performance Triboelectric Nanogenerators. *Adv. Sci.* **7**, 1–25 (2020).

33. Xu, R. *et al.* Direct current triboelectric cell by sliding an n-type semiconductor on a p-type semiconductor. *Nano Energy* **66**, 104185 (2019).

34. Bratsch, S. G. Standard Electrode Potentials and Temperature Coefficients in Water at 298.15 K. *J. Phys. Chem. Ref. Data* **18**, 1–21 (1989).

35. Turner, D. R. Junction Delineation on Silicon in Electrochemical Displacement Plating Solutions. *J. Electrochem. Soc.* **106**, 701 (1959).

36. Vivet, N. *et al.* Electrical Homo-Junction Delineation Techniques: A Comparative Study. *Mater. Sci. Appl.* **07**, 326–347 (2016).

37. Gilman, J. J. Mechanochemistry. *Science.* **274**, 65 (1996).

38. James, S. L. & Frišcic, T. Mechanochemistry. *Chem. Soc. Rev.* **42**, 7494–7496 (2013).

39.  Friščić, T., Mottillo, C. & Titi, H. M. Mechanochemistry for Synthesis. *Angew. Chemie* **132**, 1030–1041 (2020).

40.  Zi, Y. *et al.* Standards and figure-of-merits for quantifying the performance of triboelectric nanogenerators. *Nat. Commun.* **6**, (2015).

41.  Zähr, J., Ullrich, H. J., Oswald, S., Türpe, M. & Füssel, U. Analyses about the influence of the natural oxide layer of aluminium on the brazeability in a shielding gas furnace. *Weld. World* **57**, 449–455 (2013).

42.  Jeurgens, L. P. H., Sloof, W. G., Tichelaar, F. D. & Mittemeijer, E. J. Growth kinetics and mechanisms of aluminum-oxide films formed by thermal oxidation of aluminum. *J. Appl. Phys.* **92**, 1649–1656 (2002).

43.  Hunter, M. S. & Fowle, P. Natural and Thermally Formed Oxide Films on Aluminum. *J. Electrochem. Soc.* **103**, 482 (1956).

44.  Liu, J. *et al.* Sustained electron tunneling at unbiased metal-insulator-semiconductor triboelectric contacts. *Nano Energy* **48**, 320–326 (2018).

45.  Dong, S. *et al.* Freestanding-Mode Tribovoltaic Nanogenerator for Harvesting Sliding and Rotational Mechanical Energy. **2300079**, 1–8 (2023).

46.  Jeon, T. I. & Grischkowsky, D. Nature of conduction in doped silicon. *Phys. Rev. Lett.* **78**, 1106–1109 (1997).

47.  Fourmond, E. *et al.* Electrical properties of boron, phosphorus and gallium co-doped silicon. *Energy Procedia* **8**, 349–354 (2011).

48.  Kozodoy, P. *et al.* Heavy doping effects in Mg-doped GaN. *J. Appl. Phys.* **87**, 1832–1835 (2000).

# Supplementary Information

## Triboelectric Junction:
## A Model for Dynamic Metal-Semiconductor Contacts

**Xiaote Xu[1,2], Zhong Lin Wang[3,4*], Zhengbao Yang[1,2*]**

**1 Department of Mechanical and Aerospace Engineering, Hong Kong University of Science and Technology, Clear Water Bay, Hong Kong, China**
**2 Department of Mechanical Engineering, City University of Hong Kong, Hong Kong SAR, China**
**3 Beijing Institute of Nanoenergy and Nanosystems, Chinese Academy of Sciences, Beijing, 101400, People's Republic of China**
**4 School of Materials Science and Engineering, Georgia Institute of Technology, Atlanta, GA, 30332, USA**

**Email:** zbyang@ust.hk; zhong.wang@mse.gatech.edu

**This file includes:**

**Figure S1**. Metal-semiconductor contacts
**Figure S2.** Workflows of solar cell and dynamic semiconductor generator
**Figure S3.** Theoretical analysis of the triboelectric junction in dynamic metal-semiconductor contacts
**Figure S4.** Standard electrode potential
**Figure S5.** The junction direction determines the output polarity
**Figure S6.** Charge movement driven by the triboelectric junction in different contact materials
**Figure S7.** Output instability in dynamic metal-silicon contacts
**Figure S8.** Special case for dynamic metal-silicon contacts
**Figure S9.** The photovoltaic effect of intrinsic silicon-based devices
**Note S1.** The detailed derivation of the theoretical analysis
**Table S1.** Table of standard electrode potential

# Supplementary Figures

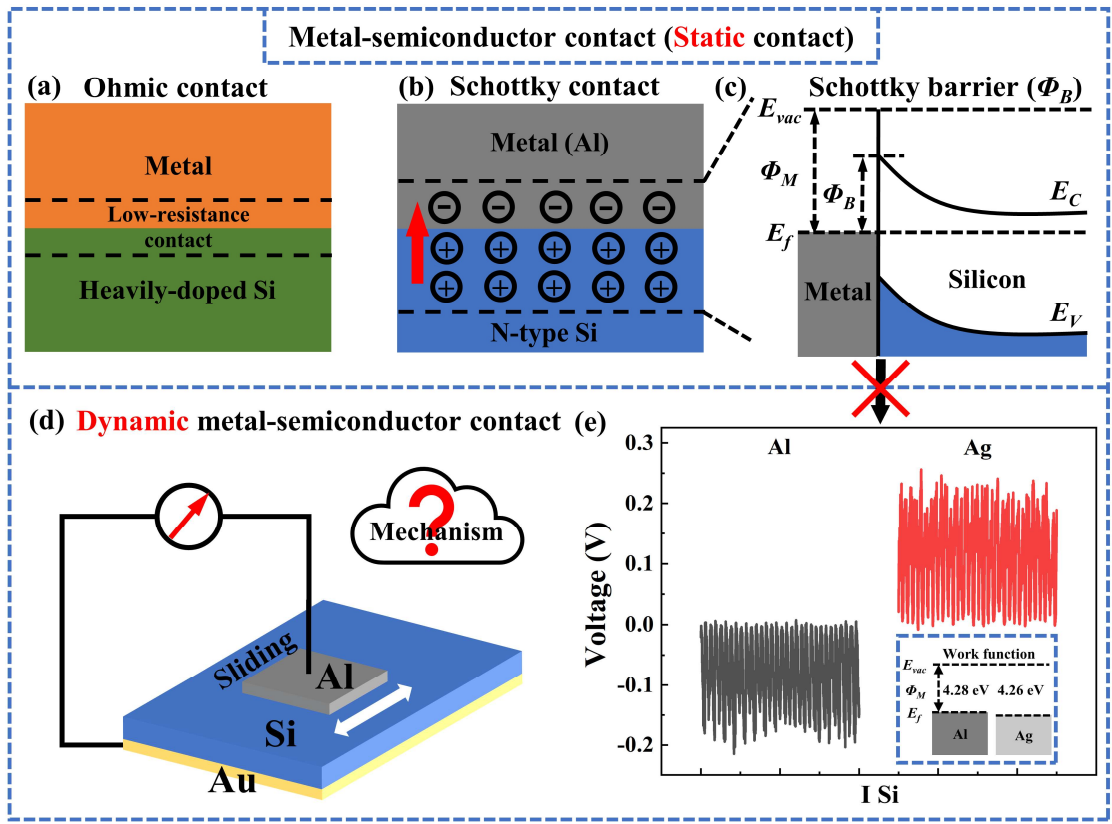

**Figure S1. Metal-semiconductor contacts.** (a) Schematic diagram of an Ohmic contact; and (B) Schottky contact. Note that the space charge region in metals is extremely thin (almost non-existent); (c) Schottky barrier model; (d) Dynamic metal-semiconductor contact; (e) Opposite output polarity of a dynamic Al-Si contact and a dynamic Ag-Si contact. Such a phenomenon in dynamic contacts cannot be explained by the classical Schottky junction theory.

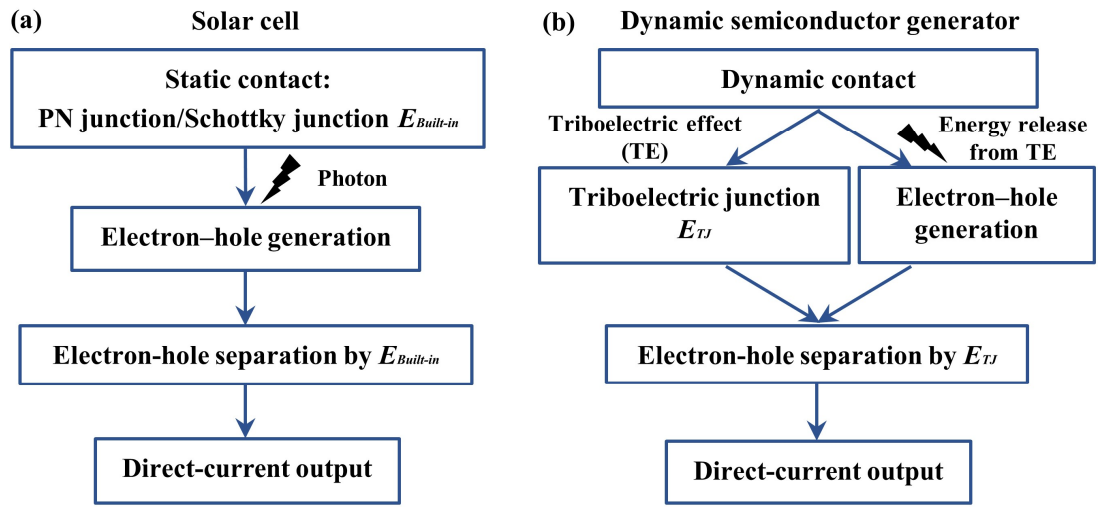

**Figure S2. Workflows of solar cell and dynamic semiconductor generator.** (a) Solar cell; and (b) Dynamic semiconductor generator.

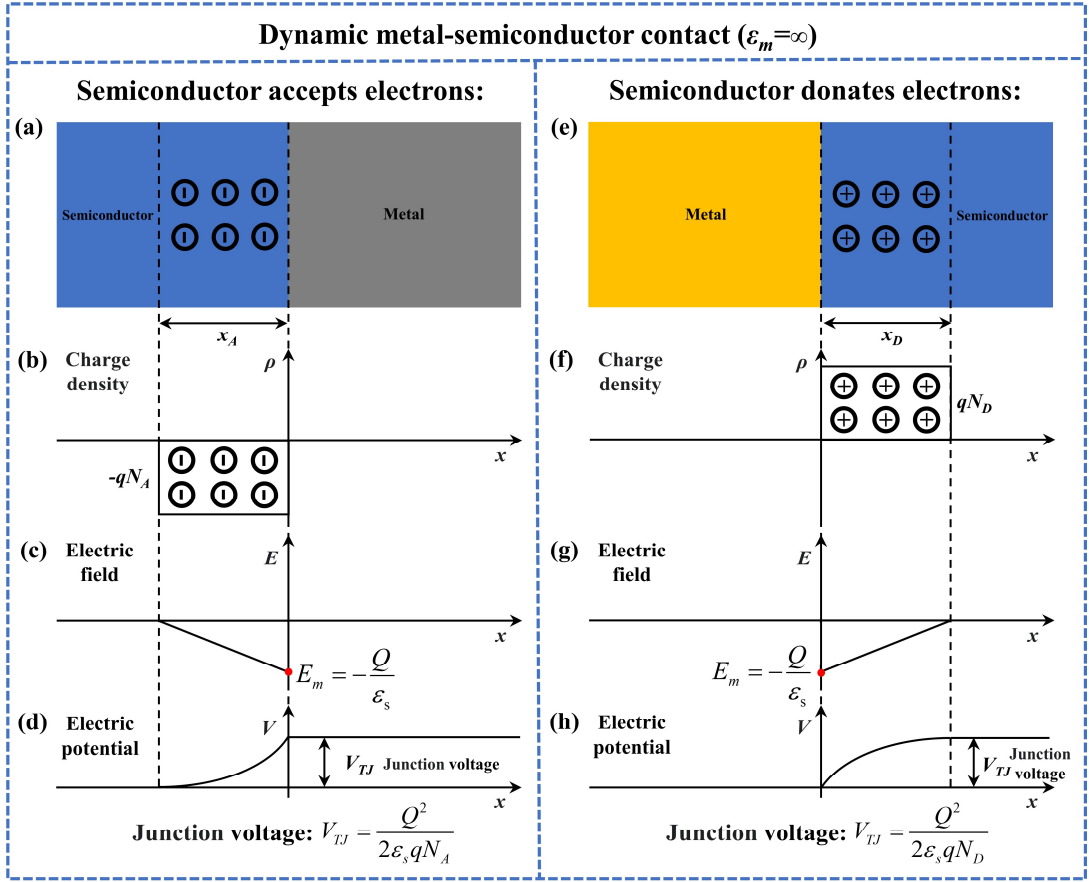

**Figure S3. Theoretical analysis of the triboelectric junction in dynamic metal-semiconductor contacts.** (a) Schematic diagram of the model that the semiconductor accepts electrons; and the corresponding (b) Charge density distribution; (c) Electric field distribution; (d) Electric potential distribution; (e) Schematic diagram of the model that the semiconductor donates electrons; and the corresponding (f) Charge density distribution; (g) Electric field distribution; (h) Electric potential distribution.

## Standard electrode potential (Unit: V)

| (-) | Mg | Al | Zn | Sn | (Si) | Cu | Ag | Pt | Au | (+) |
|---|---|---|---|---|---|---|---|---|---|---|
| Ions | $Mg^{2+}$ | $Al^{3+}$ | $Zn^{2+}$ | $Sn^{2+}$ | $SiH_4\,(g)$ | $Cu^{2+}$ | $Ag^+$ | $Pt^{2+}$ | $Au^{3+}$ | |
| φ | -2.37 | -1.66 | -0.76 | -0.13 | (0.10) | 0.34 | 0.80 | 1.19 | 1.52 | |

**Figure S4. Standard electrode potential.**

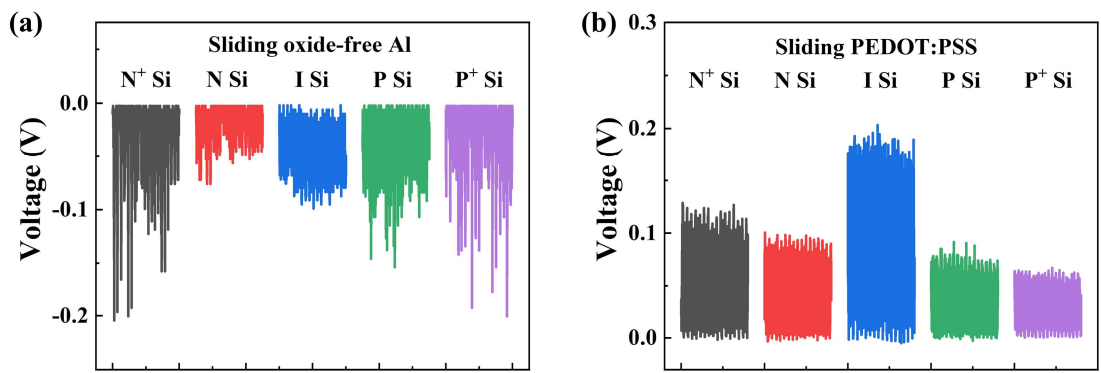

**Figure S5. The junction direction determines the output polarity.** (a) Negative outputs are observed when sliding oxide-free Al on silicon in a helium glovebox. The natural oxide layer of Al was removed by mechanically rubbing Al with sandpaper before testing; (b) Positive outputs for dynamic PEDOT:PSS-Si contacts (dynamic semiconductor-semiconductor contacts).

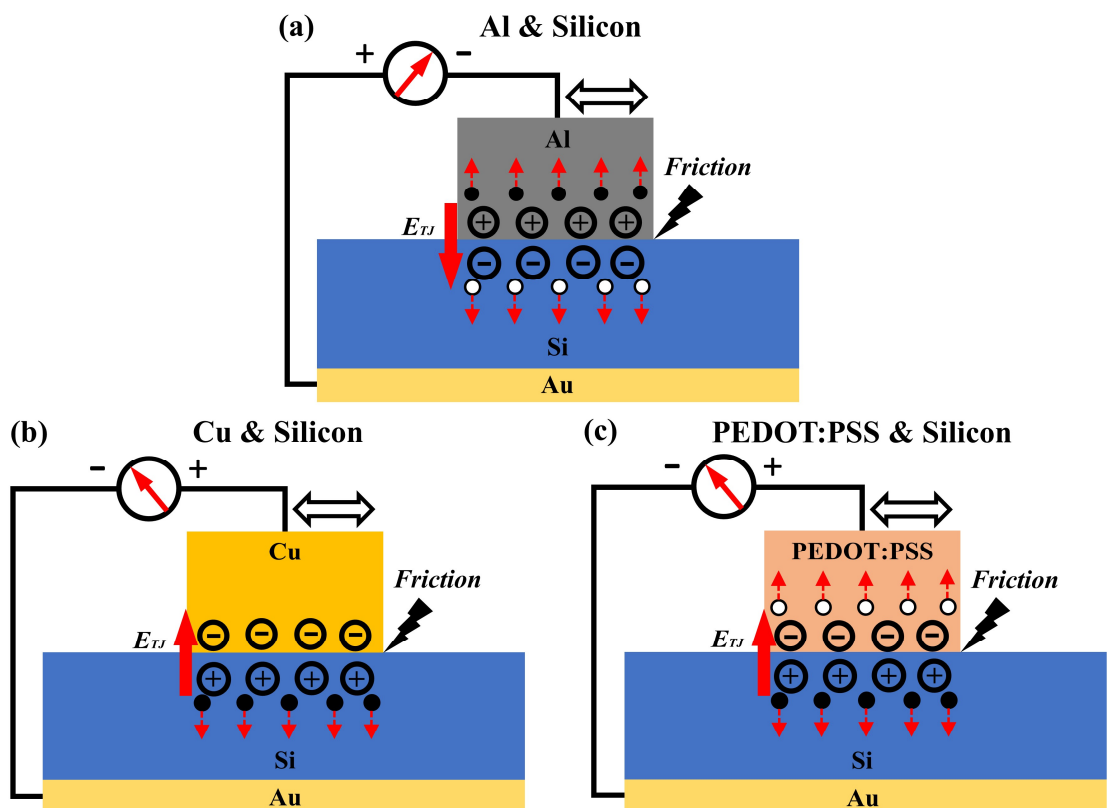

**Figure S6. Charge movement driven by the triboelectric junction in different contact materials.** (a) Dynamic Al-Si contact; (b) Dynamic Cu-Si contact. No hole transport in Cu; (c) Dynamic PEDOT:PSS-Si contact.

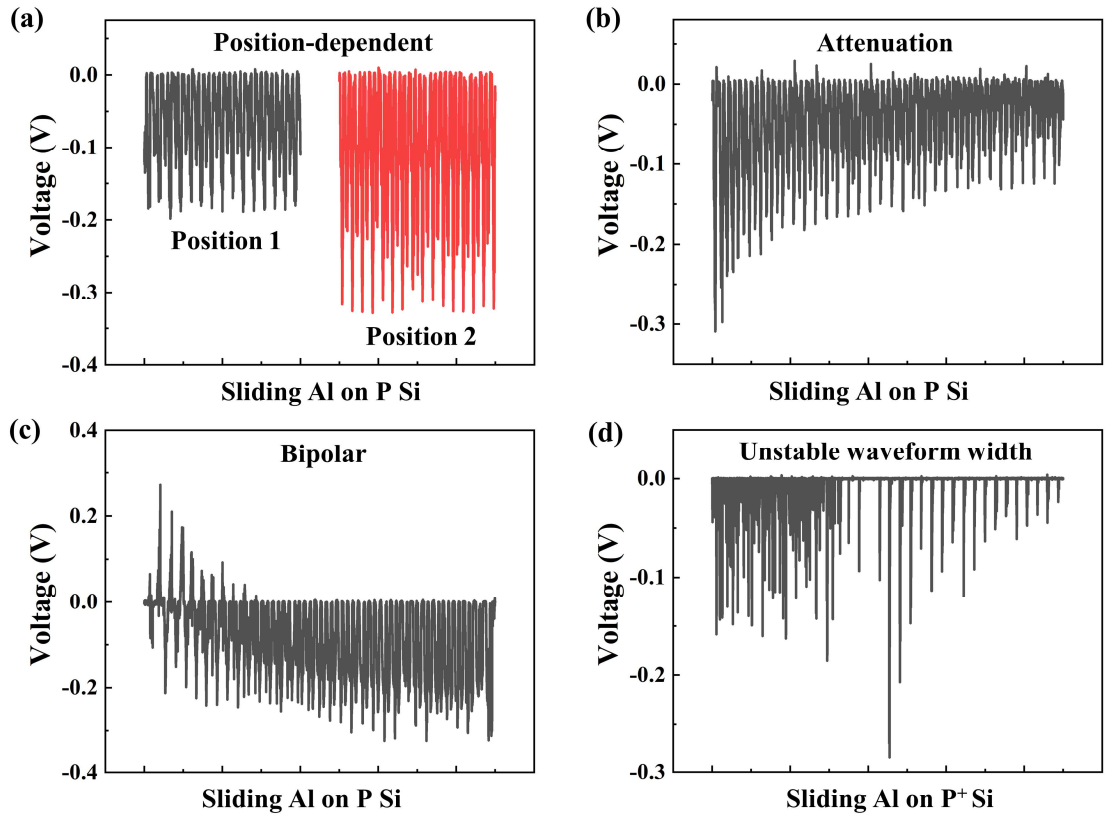

**Figure S7. Output instability in dynamic metal-silicon contacts.** (a) Position-dependent output: the output amplitude varies when sliding the same metal on different positions of silicon; (b) Output attenuation: the output voltage gradually decreases in a certain period; (c) Bipolar output: while the output polarity stabilized after a certain period, consistent with the triboelectric junction model; The above phenomena in non-degenerate silicon may be due to the surface state and/or debris induced by friction. (d) Unstable waveform width in heavily doped silicon: the waveform width varies dramatically when sliding metal on heavily doped silicon, which may be due to the strong electron-hole recombination in heavily doped silicon.

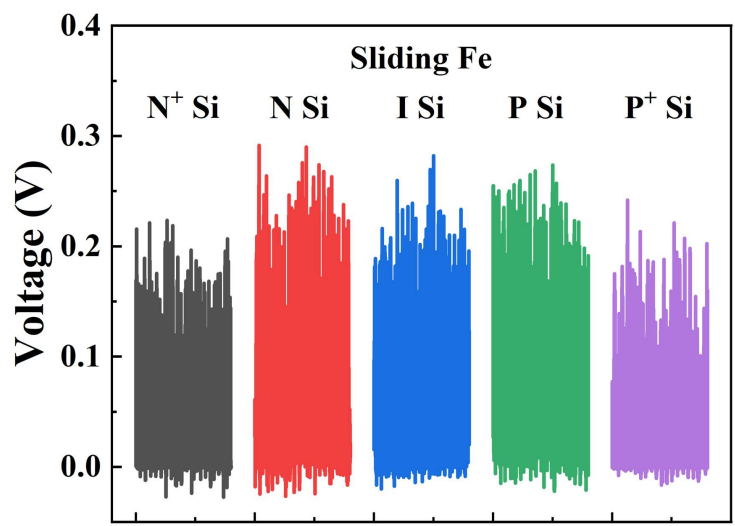

**Figure S8. Special case for dynamic metal-silicon contacts.** A dynamic Fe-Si contact generates a positive output, while a dynamic Zn-Si contact and a dynamic Sn-Si contact both generate negative outputs (Reactivity series: Zn>Fe>Sn). Fe is a metal element with multiple chemical valences. Mechanochemistry that results in another chemical valence or coexistence of multiple chemical valences, rather than the $Fe^{2+}$ used for characterizing the reactivity series, may happen in a dynamic Fe-Si contact.

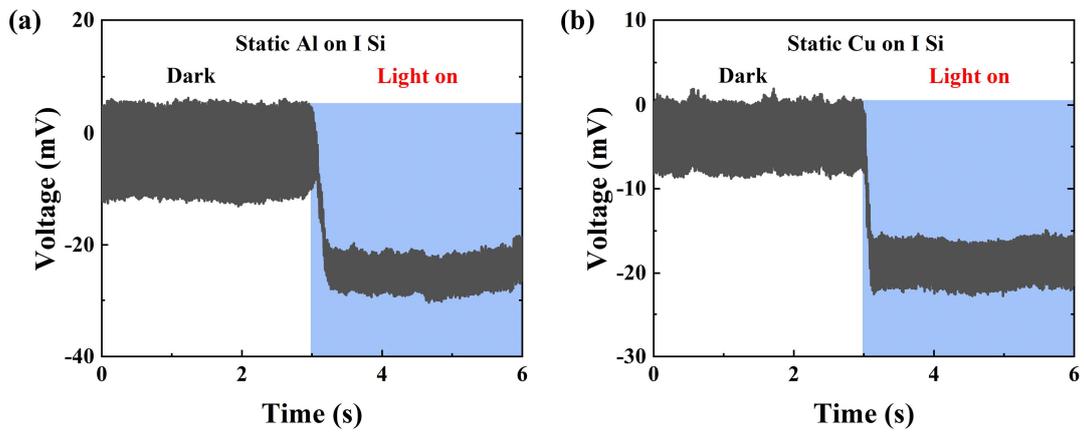

**Figure S9. The photovoltaic effect of intrinsic silicon-based devices.** (a) The photovoltaic effect of the static contact using Al and I Si (about -25 mV under ambient light in the testing lab); (b) The photovoltaic effect of the static contact using Cu and I Si (about -20 mV under ambient light in the testing lab).

# Supplementary Note

## Note S1. The detailed derivation of the theoretical analysis

A triboelectric junction is a space charge region induced by the triboelectric effect, exhibiting characteristics similar to an abrupt heterojunction. In an abrupt heterojunction, the ion types in the space charge region change abruptly from positive ions to negative ions across the contact interface. In a triboelectric junction, one material donates electrons while the other material accepts electrons, which also presents an abrupt change in ion type. Thus, the analysis of the triboelectric junction can be analogous to the abrupt heterojunction. To simplify the analysis, we discuss the triboelectric junction in a thermal equilibrium condition. In a triboelectric junction, the material that accepts (donates) electrons forms a negative (positive) ion zone with an ion concentration of $N_A$ ($N_D$) (see Fig. 2a). The widths of the negative and positive ion zone are defined as $x_A$ and $x_D$, respectively. The contact interface of the negative and positive ion zone is defined as $x=0$. Thus, we obtain the charge density distribution, as shown in Fig. 2b:

$$\rho(x) = \begin{cases} -qN_A, & -x_A \leq x \leq 0 \\ qN_D, & 0 \leq x \leq x_D \end{cases}. \tag{S1}$$

The width of the triboelectric junction:

$$X = x_A + x_D. \tag{S2}$$

According to the electron transfer model of the triboelectric effect, the space charge amount (the triboelectric charge amount) in the negative ion zone is equal to that of the positive ion zone:

$$qN_A x_A = qN_D x_D = Q, \tag{S3}$$

where $Q$ is the space charge amount (the triboelectric charge amount) per unit area.

Equation S3 can be further simplified as

$$N_A x_A = N_D x_D. \tag{S4}$$

According to the Poisson equation, we obtain

$$-\frac{d^2V(x)}{dx^2} = \frac{dE(x)}{dx} = \frac{\rho(x)}{\varepsilon(x)} = \begin{cases} -\dfrac{qN_A}{\varepsilon_A}, & -x_A \leq x \leq 0 \\ \dfrac{qN_D}{\varepsilon_D}, & 0 \leq x \leq x_D \end{cases}. \tag{S5}$$

Integrating the above equations, we obtain

$$\frac{dV(x)}{dx} = \begin{cases} (\dfrac{qN_A}{\varepsilon_A})x + C_1, & -x_A \leq x \leq 0 \\ -(\dfrac{qN_D}{\varepsilon_D})x + C_2, & 0 \leq x \leq x_D \end{cases}, \tag{S6}$$

where $C_1$ and $C_2$ are constant, which can be derived from the boundary conditions. We discuss the triboelectric junction in a thermal equilibrium condition, in which the region outside the triboelectric junction is electrically neutral. The electric field is concentrated in the triboelectric junction, so we obtain the boundary conditions:

$$\begin{cases} E(-x_A) = -\dfrac{dV(x)}{dx}\bigg|_{x=-x_A} = 0 \\ E(x_D) = -\dfrac{dV(x)}{dx}\bigg|_{x=x_D} = 0 \end{cases}. \tag{S7}$$

Substituting Equation S7 into Equation S6, we obtain

$$C_1 = \frac{qN_A x_A}{\varepsilon_A}, \quad C_2 = \frac{qN_D x_D}{\varepsilon_D}. \tag{S8}$$

Thus, we obtain the electric field distribution of the triboelectric junction, as shown in Fig. 2c:

$$E(x) = -\frac{dV(x)}{dx} = \begin{cases} -\dfrac{qN_A(x+x_A)}{\varepsilon_A}, & -x_A \le x \le 0 \\ \dfrac{qN_D(x-x_D)}{\varepsilon_D}, & 0 \le x \le x_D \end{cases}. \tag{S9}$$

The electric field strength is a linear function of the position $x$ in the negative and positive ion zone, respectively. The electric field direction is from the positive ion zone to the negative ion zone, along the negative direction of the *x-axis*. At $x$=0, the electric field strength reaches its maximum for the negative ion zone ($E_{Am}$) and positive ion zone ($E_{Dm}$), respectively. Thus, we obtain

$$E_{Am} = -\frac{qN_A x_A}{\varepsilon_A} = -\frac{Q}{\varepsilon_A}, \tag{S10}$$

$$E_{Dm} = -\frac{qN_D x_D}{\varepsilon_D} = -\frac{Q}{\varepsilon_D}. \tag{S11}$$

Note that the electric field distribution is discontinuous across the contact interface ($x$=0) due to the change in the dielectric permittivity. However, the electric displacement field is continuous ($\varepsilon_A E_{Am} = \varepsilon_D E_{Dm}$).

Integrating Equation S9, we obtain the electric potential distribution:

$$V(x) = \begin{cases} (\dfrac{qN_A}{2\varepsilon_A})x^2 + (\dfrac{qN_A x_A}{\varepsilon_A})x + D_1, & -x_A \le x \le 0 \\ -(\dfrac{qN_D}{2\varepsilon_D})x^2 + (\dfrac{qN_D x_D}{\varepsilon_D})x + D_2, & 0 \le x \le x_D \end{cases}, \tag{S12}$$

where $D_1$ and $D_2$ are constants, which can be derived from the boundary conditions. Let the electric potential of the neutral region (outside the triboelectric junction) of Material A (accepting electrons) be zero. We obtain the boundary condition under thermal equilibrium condition:

$$V(-x_A) = 0, \quad V(x_D) = V_{TJ}. \tag{S13}$$

Substituting Equation S13 into Equation S12, we obtain

$$D_1 = \frac{qN_A x_A{}^2}{2\varepsilon_A}, \quad D_2 = V_{TJ} - \frac{qN_D x_D{}^2}{2\varepsilon_D}. \tag{S14}$$

Substituting Equation S14 in Equation S12, we obtain the electric potential distribution:

$$V(x) = \begin{cases} \dfrac{qN_A(x^2 + x_A{}^2)}{2\varepsilon_A} + \dfrac{qN_A x x_A}{\varepsilon_A}, & -x_A \leq x \leq 0 \\[3mm] V_{TJ} - \dfrac{qN_D(x^2 + x_D{}^2)}{2\varepsilon_D} + \dfrac{qN_D x x_D}{\varepsilon_D}, & 0 \leq x \leq x_D \end{cases}. \tag{S15}$$

Note the electric potential distribution is continuous regardless of whether the electric field is continuous or not. At $x=0$, the electric potential is continuous:

$$V_A(0) = V_D(0) \ , \tag{S16}$$

where $V_A(x)$ and $V_D(x)$ are the electric potential distribution in the negative and positive ion zone, respectively.

Substituting Equation S16 into Equation S15, we obtain the electric potential across the triboelectric junction, namely the junction voltage:

$$V_{TJ} = \frac{qN_A x_A{}^2}{2\varepsilon_A} + \frac{qN_D x_D{}^2}{2\varepsilon_D}. \tag{S17}$$

Substituting Equation S17 into Equation S15, we obtain another expression of the electric potential distribution, as shown in Fig. 2d:

$$V(x) = \begin{cases} \dfrac{qN_A(x + x_A)^2}{2\varepsilon_A} \ , & -x_A \leq x \leq 0 \\[3mm] \dfrac{qN_D}{\varepsilon_D}(x_D - \dfrac{x}{2})x + \dfrac{qN_A x_A{}^2}{2\varepsilon_A}, & 0 \leq x \leq x_D \end{cases}. \tag{S18}$$

In the triboelectric junction, the electric potential distribution is parabolic in the negative and positive ion zone, respectively.

Substituting Equation S3 ($qN_A X_A = qN_D X_D = Q$) into Equation S17, we obtain

$$V_{TJ} = \frac{Q^2}{2\varepsilon_A q N_A} + \frac{Q^2}{2\varepsilon_D q N_D}. \tag{S19}$$

For dynamic metal-semiconductor contacts, the junction region is considered to exist only in the semiconductor ($\varepsilon_m = \infty$), similar to the one-sided abrupt PN junction (P$^+$N junction or PN$^+$ junction). The triboelectric junction of metal-semiconductor contacts can be divided into two categories and further simplified.

1) The semiconductor accepts electrons (see Fig. 2e, Fig. 2f, and Fig. S3):

The charge density distribution:

$$\rho(x) = -qN_A, \quad -x_A \leq x \leq 0 \ . \tag{S20}$$

The electric field distribution:

$$E(x) = -\frac{dV(x)}{dx} = -\frac{qN_A(x+x_A)}{\varepsilon_A}, \quad -x_A \le x \le 0 \;. \tag{S21}$$

The maximum electric field strength:

$$E_m = -\frac{Q}{\varepsilon_S} \;. \tag{S22}$$

The electric potential distribution (Let the electric potential of the neutral region be zero):

$$V(x) = \frac{qN_A(x+x_A)^2}{2\varepsilon_A} , \quad -x_A \le x \le 0 \;. \tag{S23}$$

The junction voltage:

$$V_{TJ} = \frac{qN_A x_A^{\,2}}{2\varepsilon_A} = \frac{Q^2}{2\varepsilon_S qN_A} \;. \tag{S24}$$

2) The semiconductor donates electrons (see Fig. 2g, Fig. 2h, and Fig. S3):

The charge density distribution:

$$\rho(x) = qN_D \;, \quad 0 \le x \le x_D \;. \tag{S25}$$

The electric field distribution:

$$E(x) = -\frac{dV(x)}{dx} = \frac{qN_D(x-x_D)}{\varepsilon_D}, \quad 0 \le x \le x_D \;. \tag{S26}$$

The maximum electric field strength:

$$E_m = -\frac{Q}{\varepsilon_S} \;. \tag{S27}$$

The electric potential distribution (Let the electric potential at the contact interface be zero):

$$V(x) = \frac{qN_D}{\varepsilon_D}(x_D - \frac{x}{2})x, \quad 0 \le x \le x_D \;. \tag{S28}$$

The junction voltage:

$$V_{TJ} = \frac{qN_D x_D^{\,2}}{2\varepsilon_D} = \frac{Q^2}{2\varepsilon_S qN_D} \;. \tag{S29}$$

Although both the electric field distributions and the electric potential distributions are different between these two categories, the expression of the maximum electric field strength is the same:

$$E_m = -\frac{Q}{\varepsilon_S} \;. \tag{S30}$$

The junction voltage can be concluded into the same expression:

$$V_{TJ} = \frac{Q^2}{2\varepsilon_S qN} \;, \tag{S31}$$

where $N$ is $N_A$ or $N_D$, depending on whether the semiconductor accepts or donates electrons.

## Supplementary Table

**Table S1. Table of standard electrode potential**

| Element | Half-reaction | | | $E°$ / V | Electrons |
|---------|---------|---|-----------|---------|-----------|
| | Oxidant | ⇌ | Reductant | | |
| Mg | $Mg^{2+} + 2e^-$ | ⇌ | $Mg(s)$ | -2.372 | 2 |
| Al | $Al^{3+} + 3e^-$ | ⇌ | $Al(s)$ | -1.662 | 3 |
| Zn | $Zn^{2+} + 2e^-$ | ⇌ | $Zn(s)$ | -0.7618 | 2 |
| Sn | $Sn^{2+} + 2e^-$ | ⇌ | $Sn(s)$ | -0.13 | 2 |
| Si | $Si(s) + 4H^+ + 4e^-$ | ⇌ | $SiH_4(g)$ | +0.102 | 4 |
| Cu | $Cu^{2+} + 2e^-$ | ⇌ | $Cu(s)$ | 0.337 | 2 |
| Ag | $Ag^+ + e^-$ | ⇌ | $Ag(s)$ | 0.7996 | 1 |
| Pt | $Pt^{2+} + 2e^-$ | ⇌ | $Pt(s)$ | 1.188 | 2 |
| Au | $Au^{3+} + 3e^-$ | ⇌ | $Au(s)$ | 1.52 | 3 |

Note: The data in Table S1 and Figure S4 (except for Silicon) are extracted from:
https://en.wikipedia.org/wiki/Standard_electrode_potential_(data_page)
The data for Silicon are extracted from:
https://www.av8n.com/physics/redpot.htm